\documentclass[aps,prd,twocolumn,groupedaddress]{revtex4-1}
\usepackage{graphicx}
\include{amssym}

\begin{document}
\title{$\tau\to \nu_\tau\rho^0\pi^-$ decay in the Nambu - Jona-Lasinio model}

\author{A. A. Osipov}
\email[]{aaosipov@jinr.ru} 
\altaffiliation{}
\affiliation{Bogoliubov Laboratory of Theoretical Physics, Joint Institute for Nuclear Research, Dubna, 141980, Russia}

\begin{abstract}
Within the context of an extended Nambu - Jona-Lasinio model, we analyze the role of the axial-vector $a_1(1260)$ and $a_1(1640)$ mesons in the decay $\tau\to\nu_\tau \rho^0\pi^-$. The contributions of pseudoscalar $\pi$ and $\pi (1300)$ states are also considered. The form factors for the decay amplitude are determined in terms of the masses and widths of these states. To describe the radial excited states $\pi (1300)$ and $a_1(1640)$ we introduce two additional parameters which can be estimated theoretically, or fixed from experiment. The decay rate and $\rho\pi$ mass spectrum are calculated.
\end{abstract}

\maketitle

\section{Introduction}
Semihadronic decay modes of the tau lepton remain to present date a topic of interest to theoreticians as well as experimentalists \cite{Davier06}. One mode of particular interest is the decay $\tau \to \nu_\tau \pi^+\pi^-\pi^-$. This decay is governed by the axial-vector hadronic current $j_\mu^A$ and gives a unique possibility to scrutinize  our understanding of chiral dynamics in the energy range of $1-2\,\mbox{GeV}$, where perturbative QCD methods are not applicable. There are several resonances at these energies with quantum numbers $J^{PC}=0^{-+}, 0^{++}, 1^{++}, 2^{++}$. The nature of some of these states is not yet well understood. 

The specific mode $\tau \to \nu_\tau \rho^0\pi^- \to \nu_\tau \pi^+\pi^-\pi^-$, which is the main subject of our present investigation, is most suitable to study the role of $0^{-+}$ and $1^{++}$ states in the hadronization process. Besides the pion, these are $\pi (1300)$, $a_1(1260)$, and $a_1(1640)$ resonances. In the Nambu - Jona-Lasinio (NJL) model $a_1(1260)$ is considered to be a member of the basic axial-vector nonet, i.e. $a_1(1260)$ is a pure $q\bar q$ state, with $a_1(1640)$ being its first radial excitation. The pseudoscalar $\pi (1300)$ is the first radial excitation of the pion. One of our goals here is to clarify the role of these resonances in the $\tau\to \nu_\tau \rho^0\pi^-$ decay.

In fact, the considered $q\bar q$ picture agrees with the leading order of the $1/N_c$ expansion [at large $N_c$, where $N_c$ is the number of colors, mesons are pure $q\bar q$ states, rather than, for instance, $qq\bar q\bar q$ \cite{Hooft74,Witten79}]. Of course, a more detailed description of these states would require implementation of mixing scenarios, in which the $q\bar q$ components mix with the four-quark components. This step requires to take into account the next to leading order $1/N_c$ corrections and will not be considered here. Let us also notice that for the $a_1(1260)$ axial-vector meson, there is no established understanding whether it is a quark-antiquark state or dynamically generated hadronic molecule \cite{Oset05,Wagner08,Nawa11}. Thus, it is useful to study how far one can go with the $q\bar q$ picture of $a_1(1260)$.  

Another goal of this work is to attract the attention of experimentalists to the important information contained in the specific mode $\tau\to \nu_\tau\rho^0\pi^-\to \nu_\tau\pi^+\pi^-\pi^-$, which is shown to be sensitive only to the $a_1(1260)$ and $a_1(1640)$ contributions. The experimental data on the spectral function [see Fig.\ref{fig3}] would clarify the specific role of the $a_1(1640)$ state. It is quite difficult to study the $a_1(1260)-a_1(1640)$ interference through the fit of the $3\pi$ invariant mass spectra of the $\tau\to \nu_\tau\pi^+\pi^-\pi^-$ mode, because the  corresponding amplitude has too many parameters to fit \cite{Roig13}. The major subprocess of the channel $\tau\to \nu_\tau\rho^0\pi^-\to \nu_\tau\pi^+\pi^-\pi^-$ is the $\tau\to \nu_\tau \rho^0\pi^-$ decay. The amplitude of this three-particle decay has much less number of parameters.   

The relevant approximation to this question is the $1/N_c$ expansion which provides the solid theoretical grounds for the description of the $q\bar q$ resonance states. In accord with this idea, all $q\bar q$ meson states [including $q\bar q$-resonances] are stable, free,  and non-interacting at $N_c =\infty$. It is from the point of view of the $1/N_c$ expansion the theoretical idea about an on-shell $\rho (770)$ state makes the sense, and the $\tau\to \nu_\tau \rho^0\pi^-$ decay amplitude, at leading order, can be described by the tree Feynman diagrams. 

The $\tau\to \nu_\tau\rho^0\pi^-\to \nu_\tau\pi^+\pi^-\pi^-$ mode contains the all necessary information about the $\tau\to \nu_\tau\rho^0\pi^-$ decay, that relates our study to the experiment. For instance, the sequential decay formula \cite{Cheng10} which is correct in the narrow width approximation
\begin{eqnarray}
\label{sdf}
&& \Gamma (\tau\to \nu_\tau\rho^0\pi^-\to \nu_\tau\pi^+\pi^-\pi^-) \nonumber \\
&& =\Gamma (\tau\to \nu_\tau\rho^0\pi^-)\mbox{Br} (\rho^0\to\pi^+\pi^-)
\end{eqnarray}  
and the fact that the $\rho^0$ decays into $\pi^+\pi^-$ to hundred percent yield $\Gamma (\tau\to \nu_\tau\rho^0\pi^-)=\Gamma (\tau\to \nu_\tau\rho^0\pi^-\to \nu_\tau\pi^+\pi^-\pi^-)$. Since the latter value can be extracted from the data on $\tau\to \nu_\tau\pi^+\pi^-\pi^-$, the theoretical estimate of $\Gamma (\tau\to \nu_\tau\rho^0\pi^-)$ has a definite sense [eg. recently \cite{Meissner18}, the theoretical result for $\Gamma (\eta' \to\rho \gamma)$ has been used to quantify $\Gamma (\eta' \to\rho \gamma \to \pi^+\pi^-\gamma )$ ]. It is necessary to notice that a similar situation occurs for the $\tau\to\nu_\tau\rho^0 K^-$ decay, where the sequential decay mode $\tau\to \nu_\tau\rho^0 K^-\to \nu_\tau\pi^+\pi^- K^-$ has already been measured [the PDG quoted value is Br($\Gamma (\tau\to \nu_\tau\rho^0 K^-\to \nu_\tau\pi^+\pi^- K^-))=(1.4\pm 0.5) \times 10^{-3}$ \cite{PDG18}]. The measurement of the Br($\Gamma (\tau\to \nu_\tau\rho^0 \pi^-\to \nu_\tau\pi^+\pi^- \pi^-))$ will not only fill the gap in the existing data, but, as it is shown in this work, will clarify the role of the $a_1(1640)$ resonance in the underlying chiral dynamics.     

In the NJL model, there is a nonlocal extension which deals with the excited states of the $0^{-+}, 0^{++}, 1^{--}$ and $1^{++}$ ground state nonets \cite{Weiss97,Volkov97}. The nonlocal four-quark interactions lead to the nonlocal effective meson Lagrangian which describes the physics of these excited states. Nonetheless, here we will apply a more modest description, which should ideally arise from \cite{Weiss97,Volkov97} in the large $N_c$ limit, namely we suppose that excited states at leading $1/N_c$ order can be described by the local Lagrangian, like, for instance, in the extended linear sigma model approach \cite{Giacosa17}. In this case the propagators of excited states have the same form as the ground state propagators, but with different couplings to the weak axial-vector current. Such a simplified treatment of exited states is not new. Notably, it is exactly how the contribution to the $\tau \to \nu_\tau \pi\pi\pi$ amplitude from the vector $\rho (1450)$ resonance exchange has been estimated in \cite{Pich10}.   
  
One of the first theoretical studies of the role of the $a_1(1260)$ axial-vector state in the $\tau \to \nu_\tau \rho^0 \pi^-$ decay is presented in \cite{Geffen78}, where the current algebra sum rules have been used to clarify whether experimental data are compatible with a contribution of the $a_1(1260)$ resonance to the $\tau \to \nu_\tau \rho^0\pi^-$ decay mode. This decay has been also considered in \cite{Li97}, where the $a_1(1260)$ dominance has been revealed. There is no doubt, nowadays, about the dominant role of $a_1(1260)$ in this process. However, we still need to understand the nature and parameters of the $a_1(1260)$ resonance. Besides that, the role of its radial excited state $a_1(1640)$ must be clarified. The measurements of the branching ratio and $\rho^0 \pi^-$ mass spectrum can provide insight into the issue. 

In this paper we calculate the $\tau \to \nu_\tau \rho^0\pi^- $ decay amplitude in the framework of NJL model with $SU(2)\times SU(2)$ chiral symmetry. The first attempt to use the NJL approach to study this decay was made in \cite{Osipov89}, where the finite terms in the derivative expansion of quark loops corresponding to vertices $a_1\rho\pi$ and $\rho\pi\pi$ have been taken into account. Though the analysis in \cite{Osipov89} allows to reproduce the experimental value of the $\tau \to \nu_\tau \rho^0\pi^- $ decay width [mainly due to the contribution of finite terms] the procedure of extracting these finite terms, used in \cite{Osipov89}, is not compatible with the chiral symmetry restrictions imposed on such contributions by the chiral invariant Schwinger-DeWitt expansion at large distances [the consistent Schwinger-DeWitt approach requires also to take into account the finite terms of self-energy diagrams which will redefine the coupling constants of the theory, what has not been done there]. As opposed to this, we do not consider here the contributions of the problematic finite terms, but show instead that the decay can be described by the standard effective meson Lagrangian \cite{Volkov84,Volkov86,Ebert86} provided the first radial excitations $\pi (1300)$ and $a_1(1640)$ are taken into account. 

The material of the paper is presented in the following way. In Sec. \ref{la} we describe the relevant meson vertices of the effective Lagrangian, obtain the amplitude of the $\tau \to \nu_\tau \rho^0\pi^- $ decay, and discuss the partial conservation of the axial-vector current (PCAC). This important relation should be fulfilled in the chiral approach. In Sec. \ref{es} the radial excited states are considered. We show that the inclusion of these states can be done without contradiction with the PCAC condition. In Sec. \ref{cdw} we introduce the momentum dependent widths of the resonances and calculate the differential decay width of the process. In Sec. \ref{nr} the results of our numerical calculations are presented in Tables \ref{scase} and \ref{fcase} and Fig. \ref{fig3}. We conclude with Sec. \ref{concl}.

\section{Lagrangian and amplitude}
\label{la}
Our starting point is the effective meson Lagrangian obtained on the basis of the NJL model with the global $U(2)_R\times U(2)_L$ chiral symmetric four-quark interactions which also possesses the gauge $SU(2)_L\times U(1)_R$ symmetry of the electroweak interactions. For convenience, we refer to the paper \cite{OsipovHZ18} which contains all necessary details related with the obtention of such effective Lagrangian.

The appropriate weak hadronic part of the Lagrangian density is  
\begin{equation}
\label{wh}
{\cal L}=-G_F V_{ud} l_\mu \left( \frac{m_\rho^2}{g_\rho} a_{1\mu}^+  +f_\pi \partial_\mu \pi^+ \right) 
+ \mbox{H.c.},
\end{equation}
where $G_F$ is the Fermi constant, $V_{ud}\simeq \cos\theta_c$ is an element of the Cabibbo-Kobayashi-Maskawa matrix, $\theta_c=13^\circ$ is the Cabibbo angle, $l_\mu = \bar\nu_\tau \gamma^\mu (1-\gamma_5)\tau$ is the weak charged lepton current, $m_\rho$ is the mass of the $\rho (770)$ meson, $g_\rho$ is the coupling which arises due to a redefinition of the spin-1 fields, $f_\pi =92\,\mbox{MeV}$ is the pion decay constant. 

Notice that the Lagrangian density (\ref{wh}) has the standard form of the axial-vector dominance, i.e. it does not have a contact term $\sim\rho^0 \pi^+$. In the covariant formulation \cite{OsipovHZ18} the corresponding  part of the hadronic weak axial-vector current $j^A_\mu$ is proportional to $\pi^+ (\rho^0_\mu - \partial_\nu\rho^0_{\mu\nu}/m_\rho^2)$. This factor is zero on the mass shell of the $\rho$-meson. This does not mean that the matrix element $\langle \rho \pi |j^{A}_\mu |0\rangle$ has no contact part. Hereafter, in Eq.(\ref{ff}), it is clear that it does [see the first term $\sim g_{\mu\nu}$ ]. This term is effectively originated by the $a_1$-exchange contribution $\langle \rho \pi |a_1\rangle\langle a_1| j^{A}_\mu |0\rangle$. Thus the fact that the on-shell $\rho (770)$ cancels the direct term $\propto \rho\pi$ in $j^{A}_\mu$ is a part of the fundamental mechanism which is responsible for the PCAC relation in the model.   

The hadronic weak axial-vector current $j^A_\mu$ in formula (\ref{wh}) can be also obtained in the standard noncovariant approach \cite{Osipov17ap} by using the variational method of Gell-Mann and L\'evy \cite{Levy60,Sakurai69}. However, it requires some work, because the direct application of this technique leads only to the pion exchange. To arrive at the axial-vector dominance one should use the Lagrangian equations for the axial-vector field, and after that neglect the total derivatives of the antisymmetric tensors. Another way to obtain the Lagrangian density (\ref{wh}) is described in \cite{Volkov91}.        

Thus we need only two additional vertices to find the amplitude of the $\tau \to \nu_\tau \rho^0\pi^- $ decay. This is the $a_1\rho\pi$ vertex
\begin{eqnarray} 
\label{arp}
\mathcal L_{a_1\rho\pi}&=&\frac{i}{4} Zf_\pi g^2_\rho\, \mbox{tr} \left\{ a_\mu [\rho_\mu , \pi  ] \right.\nonumber \\
&+&\left. \frac{1}{m_{a_1}^2}\left( \rho_{\mu\nu} [a_\mu ,\partial_\nu\pi ]+ a_{\mu\nu} [\rho_\mu ,\partial_\nu\pi ]\right)\right\},
\end{eqnarray}
where $\rho_{\mu\nu}=\partial_\mu\rho_\nu -\partial_\nu\rho_\mu$, $Z=(1-6m^2/m^2_{a_1})^{-1}$, $m$ is the mass of the constituent quark; $m_{a_1}=\sqrt{Z}m_\rho$ is the mass of the $a_1(1260)$ meson; it is assumed that all fields are contracted with Pauli matrices, for instance, $\pi=\pi_i\tau_i$, and trace is calculated over products of tau-matrices. Notice that the Lagrangian density (\ref{arp}) obtained in the covariant approach \cite{OsipovHZ18} coincides with the result of the standard noncovariant approach \cite{Osipov17ap}.    

The second vertex that we need, in the covariant approach, is given by the Lagrangian density 
\begin{equation}
\label{rpp}
\mathcal L_{\rho\pi\pi}=\frac{-ig_\rho}{4m_\rho^2}\left(\frac{Z+1}{Z}\right)\,\mbox{tr}\left\{\partial_\mu\rho_\nu [\partial_\mu\pi, \partial_\nu\pi ]\right\}. 
\end{equation}
On the mass shell of the $\rho$-meson it yields
 \begin{equation}
\label{rppms}
\mathcal L_{\rho\pi\pi}^{\rho-mass}=\frac{ig_\rho}{8}\left(\frac{Z+1}{Z}\right)\,\mbox{tr}\left\{\rho_\mu [\partial_\mu \pi, \pi ]\right\}. 
\end{equation}

\begin{figure}
\resizebox{0.25\textwidth}{!}{%
 \includegraphics{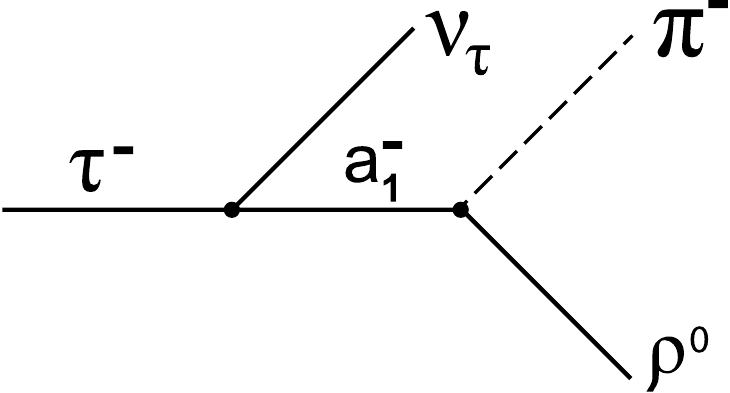}}
\caption{The Feynman diagram describing the contribution of the mode $\tau^-\to \nu_\tau a^-_1 \to \nu_\tau \rho^{0} \pi^{-}$ to the decay amplitude (\ref{ampl}).}
\label{fig1}      
\end{figure}

\begin{figure}
\resizebox{0.25\textwidth}{!}{%
 \includegraphics{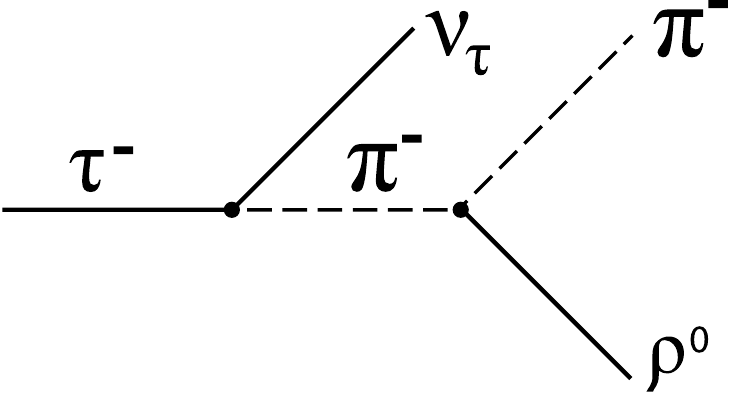}}
\caption{The Feynman diagram describing the contribution of the mode $\tau^-\to \nu_\tau \pi^- \to \nu_\tau \rho^{0} \pi^{-}$ to the decay amplitude (\ref{ampl}).}
\label{fig2}      
\end{figure}

Now we have all necessary ingredients to find the amplitude $A$ of the $\tau (Q)\to \nu_\tau (Q')+\rho^0 (p) +\pi^-(q) $ decay, where $Q,Q',p$ and $q$ are the 4-momenta of the particles. The corresponding Feynman diagrams are shown in Figs. \ref{fig1} and \ref{fig2}.

In this way we have 
\begin{equation}
\label{ampl}
A=-iG_FV_{ud} g_\rho f_\pi \bar\nu_\tau (Q') \gamma^\mu (1-\gamma_5) \tau (Q) F_\mu ,
\end{equation}
where the pure hadronic part is given by the 4-vector
\begin{equation}
\label{ff}
F_\mu=\left[g_{\mu\nu} + m^2_\rho \frac{g_{\mu\nu} -\frac{k_\mu k_\nu}{m^2_{a_1}}}{m_{a_1}^2-k^2}+ 
\left(1+\frac{1}{Z}\right)\frac{k_\mu k_\nu}{m_\pi^2-k^2}\right]\epsilon^*_\nu (p).
\end{equation}
Here $\epsilon_\nu (p)$ is a polarization vector of the $\rho$-meson, and $k=q+p$. The invariant subsidiary condition on the components of the vector state is assumed $p_\nu\epsilon_\nu (p)=0$. 

Notice that
\begin{eqnarray} 
\label{pcac}
k_\mu F_\mu &=& \left(1+\frac{m_\rho^2}{m_{a_1}^2} + \frac{Z+1}{Z} \frac{k^2}{m^2_\pi-k^2} \right) k_\nu \epsilon^*_\nu (p)
\nonumber \\
&=& \frac{Z+1}{Z} \frac{m_\pi^2}{m^2_\pi-k^2}k_\nu \epsilon^*_\nu (p),
\end{eqnarray}
i.e. $k_\mu F_\mu =\langle \pi\rho |\partial_\mu j^A_\mu|0\rangle$ is dominated by the pion pole in accord with PCAC.

We further stress the presence of a contact contribution in (\ref{ff}). The first term with $g_{\mu\nu}$ results from the diagram of Fig.\ref{fig1}. Its appearance is partly due to the first term of the Lagrangian density (\ref{arp}).    

The most general Lorentz-covariant form of $F_\mu$ is
\begin{equation}
\label{gff}
F_\mu=  \epsilon^*_\mu (p) F_0 +(p+q)_\mu (\epsilon^* q) F_- +(p-q)_\mu (\epsilon^* q) F_+.
\end{equation}
In particular, the NJL model yields 
\begin{eqnarray}
\label{aff}
F_0&=& 1+\frac{m_\rho^2}{m^2_{a_1}-k^2}, \quad F_+=0, \nonumber\\
F_-&=& \frac{1+g_A}{m^2_\pi-k^2} - \frac{g_A}{m^2_{a_1}-k^2}, 
\end{eqnarray}
where $g_A=1/Z$. Thus, the NJL approach is quite restrictive: the form factor $F_+$ does not contribute at leading order in $1/N_c$ and derivative expansions. All form factors are the functions of only one variable $k^2$.

\section{Radially Excited States and PCAC}
\label{es}
The region between $1.2-1.8\, \mbox{GeV}$ of the $\rho^0\pi^-$ spectrum is still poorly described by the standard NJL model. One can improve our description of the $\tau \to \nu_\tau \rho^0\pi^- $ decay amplitude by including the contributions of the radially excited states of the pion and the $a_1(1260)$ meson, i.e. the $\pi'=\pi (1300)$ and $a'_1=a_1(1640)$ resonances. Following \cite{Pich10} we perform the substitutions in the pion and $a_1(1260)$ propagators: 
\begin{eqnarray}  
\label{sub}
\frac{1}{m_\pi^2-k^2}&\to&\frac{1}{1+\beta_{\pi'}}\left(
\frac{1}{m_\pi^2-k^2}+\frac{\beta_{\pi'}}{m_{\pi'}^2-k^2}\right), \nonumber\\
\frac{1}{m_{a_1}^2-k^2}&\to&\frac{1}{1+\beta_{a'_1}}\left(
\frac{1}{m_{a_1}^2-k^2}+\frac{\beta_{a'_1}}{m_{a'_1}^2-k^2}\right).  
\end{eqnarray}
Notice that the limit $\beta\to 0$ corresponds to the case without excitations; other limits $m_{\pi'}\to m_{\pi}$ and $m_{a'_1}\to m_{a_1}$ lead to the same result. The substitutions are written in terms of physical states, therefore, the coupling $\beta$ is the only parameter which absorbs contributions arising due to the redefinition of the primary meson fields [this includes the diagonalization of $\pi -\pi'$ and $a_1-a_1'$ quadratic forms and the pseudoscalar - axial-vector mixing effects]. As a result, the factor $1/(1+\beta)$ rescales the contribution of the ground state. 

Doing these replacements, we must ensure that substitutions (\ref{sub}) do not destroy the PCAC condition. For that, together with (\ref{sub}), one should modify the contact term
\begin{equation}
g_{\mu\nu}\to (1+\delta )g_{\mu\nu},
\end{equation}
where $\delta$ is the constant approximating the higher-mass $1^{++}$ contribution to $F_0$ in such a way that the PCAC condition is fulfilled.

Indeed, in this case the modified hadronic part of the amplitude $A\to A'$, $F_\mu\to F_\mu'$ has the form   
\begin{eqnarray}
\label{mff}
F'_\mu&=&\epsilon^*_\nu (p) \left[(1+\delta)g_{\mu\nu} \right. \nonumber \\
&+&\frac{m^2_\rho}{1+\beta_{a'_1}} \left( 
\frac{g_{\mu\nu} -\frac{k_\mu k_\nu}{m^2_{a_1}}}{m_{a_1}^2-k^2}+\beta_{a'_1} \frac{g_{\mu\nu} -\frac{k_\mu k_\nu}{m^2_{a'_1}}}{m_{a'_1}^2-k^2} \right)  \nonumber \\
&+& \left. \frac{1+g_A}{1+\beta_{\pi'}} k_\mu k_\nu \left(\frac{1}{m_\pi^2-k^2}+\frac{\beta_{\pi'}}{m_{\pi'}^2-k^2}\right) \right].
\end{eqnarray}
The divergence of the hadronic current $k_\mu F'_\mu$ should vanish in the chiral limit $m_\pi\to 0$. This requirement is fulfilled only if $\delta$ is uniquely fixed as
\begin{equation}
\delta = \frac{g_A \beta_{a'_1}}{1+\beta_{a'_1}}  \left(1-\frac{m_{a_1}^2}{m_{a'_1}^2}\right).
\end{equation}     
As a result we obtain that
\begin{equation}
k_\mu F'_\mu=\frac{1+g_A}{1+\beta_{\pi'}} \left(\frac{m_\pi^2}{m^2_\pi-k^2} + \frac{\beta_{\pi'} m_{\pi'}^2 }{m^2_{\pi'}-k^2}\right) (\epsilon^* k).
\end{equation}
This is a modified PCAC relation, which, in particular, tells us that in the chiral limit $\beta_{\pi'}=f_{\pi'} /f_\pi \to 0$, where $f_\pi$ and $f_{\pi'}$ are the weak decay constants of the $\pi$ and $\pi'$ mesons.

Thus, the consideration above shows that to lowest order in $1/N_c$ our procedure introduces only four additional parameters: the two masses of $\pi (1300)$ and $a_1(1640)$ resonances and two mixing parameters $\beta_{\pi'}$ and $\beta_{a'_1}$ which should be fixed theoretically or from experimental data.

\section{Decay width} 
\label{cdw}
Let us proceed now with calculations of the decay rate. For that we need the appropriate spin-averaged matrix element squared  
\begin{equation}
\label{ms}
|A'|^2=(2G_FV_{ud}g_\rho f_\pi )^2 \left[2(QF')(Q'F')-(F'F')(QQ') \right]. 
\end{equation}
Here we averaged over initial $\tau$-lepton states. It is easiest to perform the calculation in invariant form before specializing to the rest frame of tau. The invariant Mandelstam variables are $s=(Q-Q')^2=k^2$, $t=(Q-q)^2$, and $u=(Q-p)^2$. It gives
\begin{eqnarray}
&&|A'|^2=\frac{G_F^2}{2}V_{ud}^2(1-g_A) \left\{             
|F'_-|^2 m_\tau^2 (m_\tau^2-s) \lambda(s,m^2_\rho,m^2_\pi )  \right. \nonumber\\
&&+4|F'_0|^2\left[t^2+t(s-m^2_\pi-m^2_\rho)+m^2_\rho (m_\tau^2-2s+m^2_\pi ) \right] \nonumber \\
&&+2\mbox{Re}(F'_0F'^*_-) \left[(m_\tau^2-s)\left((s-m^2_\pi)^2 -\lambda (s,m^2_\rho,m^2_\pi )  \right) \right. \nonumber\\
&&+m_\tau^2\left(2t(s+m_\rho^2-m_\pi^2) -m_\rho^2 (m_\rho^2+4m_\tau^2) \right) \nonumber \\
&&  +\left.\left.  m_\rho^2 s\left(2(s+m_\pi^2)-m_\rho^2\right)\right] \right\},
\end{eqnarray}
where the NJL model relation $(1-g_A)m_\rho^2 = g_\rho^2 f_\pi^2$ has been used, and the K\"all\'en function $\lambda (x,y,z)$ is defined as follows $\lambda (x,y,z)=(x-y-z)^2 - 4yz$.

In the physical range $(m_\rho+m_\pi)\leq \sqrt{s}\leq m_\tau$ the form factors in Eq. (\ref{mff}) contain zero-width $a_1$, $a'_1$ and $\pi'$ propagator poles, which lead to divergent phase-space integrals in the calculation of the $\tau\to\nu_\tau\rho^0\pi^-$ decay width. In order to regularize the integrals one should include the finite widths of these resonances through the typical Breit-Wigner form of propagators. This is a step beyond the leading order in the $1/N_c$ expansion which we are forced to make in connection with the above-mentioned problem. We consider the substitutions:
\begin{equation}
\label{Rwidth}
\frac{1}{m_R^2-k^2} \to \frac{1}{m_R^2-k^2 -im_R \Gamma_R(k^2)}.
\end{equation}
A $k^2$ dependence for $\Gamma_R(k^2)$ is required by unitarity. The description of a set of resonances with the same quantum numbers as a sum of Breit-Wigner amplitudes may violate unitarity and is a good approximation only for well-separated resonances with little overlap. This condition is fulfilled here.

Following \cite{Geffen78}, we have chosen to use the form
\begin{equation}
\label{width-1}
\Gamma_R(k^2)=\Gamma_R \sqrt{\frac{\lambda(k^2,m_\rho^2,m_\pi^2)}{\lambda(m_{R}^2,m_\rho^2,m_\pi^2)}}\frac{m_R^2+k_R^2}{k^2+k_R^2},
\end{equation}
where $\Gamma_R=\Gamma_R(m_R^2)$. The function $\Gamma_R(k^2)$ has a threshold factor in the proper position, i.e. at $k^2 =(m_\rho + m_\pi )^2$. The value of $k_R^2$ is determined by the following integral condition
\begin{equation}
\label{norm}
\frac{1}{\pi}\!\!\!\!\!\int\limits_{(m_\rho + m_\pi )^2}^\infty \!\!\!\!\!dk^2\, \mbox{Im}\left[m_R^2 -k^2 -im_R \Gamma_R(k^2)\right]^{-1}=1.
\end{equation}
In the narrow-width approximation this equation is automatically fulfilled. If the resonance is broad, Eqs. (\ref{width-1}) and (\ref{norm}) make our results less sensitive to the details of $\Gamma_R(k^2)$. A rigorous form can only be obtained if the total width is completely understood, but this is not the case at the moment. 

The differential decay rate can be written in the form 
\begin{equation}
\label{dr1}
\frac{d\Gamma}{ds}=\int_{t_-(s)}^{t_+(s)} dt\, \frac{|A'|^2}{32m_\tau^3(2\pi)^3},
\end{equation}
where 
\begin{eqnarray}
t_\pm (s)&=&\frac{1}{2}\left[m_\tau^2+m_\rho^2+m_\pi^2 -s +\frac{m_\tau^2}{s}(m_\rho^2-m_\pi^2) \right.\nonumber\\
&\pm&\left. \sqrt{D(s)} \right], \\
D(s)&=& \left(\frac{m^2_\tau}{s}-1\right)^2\lambda (s,m^2_\rho,m^2_\pi ). 
\end{eqnarray}

The integral over $t$ in (\ref{dr1}) can be done explicitly:
\begin{eqnarray}
\label{spf}
\frac{d\Gamma}{ds}&=&\frac{(G_FV_{ud})^2}{64m_\tau^3(2\pi)^3}\sqrt{D(s)} \left(\frac{m^2_\tau}{s}-1\right) (1-g_A)\nonumber\\
&\times& \left\{ 4m_\rho^2 |F'_0|^2\left[\frac{s+2m_\tau^2}{6sm_\rho^2} \lambda (s,m^2_\rho,m^2_\pi )+m_\tau^2+2s \right]\right.\nonumber \\
&+&\left. m_\tau^2 \lambda (s,m^2_\rho,m^2_\pi ) \left(s|F'_-|^2+ 2\mbox{Re}(F'_0F'^*_-)  \right)\right\}.
\end{eqnarray} 
Integrating this expression over $s$ one finally obtains the $\tau\to\nu_\tau\rho^0\pi^-$ decay width.

\section{Numerical Results}
\label{nr}
Our consideration above shows that we are able to describe the tree $\tau\to\nu_\tau\rho^0\pi^-$ decay amplitude in terms of the known masses: $m_\pi, m_\rho, m_{a_1}, m_{\pi'}, m_{a'_1}, m_\tau$, two mixing parameters $\beta_{\pi'}, \beta_{a'_1}$, and three widths $\Gamma_{a_1}, \Gamma_{a'_1}, \Gamma_{\pi'}$. The value of $g_A$ is not free due to the mass formula $g_A m_{a_1}^2=m_\rho^2$ which is valid in the NJL model. This parameter is also related with the value of the constituent quark mass $m$, which, in the case of exact isospin symmetry $m= m_u= m_d$, is given by  
\begin{equation}
m=\frac{m_\rho}{\sqrt{6}}\sqrt{\frac{1}{g_A}-1}.
\end{equation}
In the following we will vary the value of $m_{a_1}$ in the interval $1120\,\mbox{MeV} \leq m_{a_1} \leq 1300\,\mbox{MeV}$. The parameter $g_A$ will be changed correspondingly. The upper boundary is inspired by the recent measurements of the COMPASS Collaboration $m_{a_1}=1299^{+12}_{-28}\,\mbox{MeV}$, $\Gamma_{a_1}=380\pm 80\,\mbox{MeV}$ \cite{Wallner18,Aghasyan18}. The lower boundary is a result of a comparison between the theoretical $m_{3\pi}^2$-spectra of the $\tau\to\nu_\tau\pi^+\pi^-\pi^-$ decay \cite{Pich10} with ALEPH data \cite{Barate98}, that yields $m_{a_1} = 1120\,\mbox{MeV}$ and $\Gamma_{a_1}=483\pm 80\,\mbox{MeV}$. The PDG averaged values: $m_{a_1}=1230\pm 40\,\mbox{MeV}$, $\Gamma_{a_1}=250-600\,\mbox{MeV}$ \cite{PDG18}, and the parameters extracted by the JPAC group $m_{a_1}=1209\pm 4^{+12}_{-9}\,\mbox{MeV}$, and $\Gamma_{a_1}=576\pm 11^{+80}_{-20}\,\mbox{MeV}$ \cite{Mikhasenko18} are also considered.

\subsection{Ground states contribution}
Our numerical calculations we start from the simplest case, when only ground states are considered. With this purpose we use the form factors given by Eq. (\ref{aff}) modified by the substitutions (\ref{Rwidth}). As can be seen from Table \ref{scase}, the higher value of $m_{a_1}$ and the lower value of $\Gamma_{a_1}$, the better agreement with experimental data. We make this conclusion by confronting our results to the old value of the branching ratio Br$(\tau\to\nu_\tau \rho^0 \pi^-) =(5.4\pm1.7)\%$, quoted by PDG \cite{PDG88} [presently, PDG does not give data on the $\tau\to \nu_\tau \rho^0\pi^-$ mode]. This corresponds to the following decay width 
\begin{equation}
\label{exp}
\Gamma^{exp}_{\tau\to\nu_\tau\rho^0\pi^-}=\left(1.22\pm 0.39 \right) \times 10^{-10}\,\mbox{MeV}.
\end{equation}

It is worth pointing out that the latest measurements of the CLEO Collaboration of the $\tau\to\nu_\tau \pi\pi\pi$ decay \cite{Asner99,Briere03} and the results of the COMPASS experiment in diffractive production \cite{Adolph17,Wallner18,Aghasyan18} can be used [although this is a model dependent procedure which is also influenced by the production mechanism] to extract the branching ratios of the specific decay channels. In particular, the JPAC \cite{Mikhasenko18} made a rough estimate for the dominant $\rho\pi$ $S$-wave channel. The found branching ratio is of 60\% -- 80\%. It corresponds to the decay width 
\begin{equation}
\label{JPAC}
\Gamma_{\tau\to \nu_\tau \rho^0\pi^-}^{JPAC}=\left(1.26 - 1.69 \right) \times 10^{-10}\,\mbox{MeV}.
\end{equation}

\begin{table}[tb]
\caption{The width of the $\tau\to\nu_\tau\rho^0\pi^-$ decay obtained in the NJL model with only the ground state contributions. The first two columns contain the phenomenological input values of $m_{a_1}$ and $\Gamma_{a_1}$ taken from \cite{PDG18,Wallner18,Aghasyan18,Barate98,Mikhasenko18}.}
\label{scase}
\begin{center}
\begin{tabular}{cccc}
\hline
\hline
set &
    $m_{a_1}$ (MeV) & 
    $\Gamma_{a_1}(m_{a_1}^2)$ (MeV)& 
    $\Gamma_{\tau\to\nu_\tau\rho^0\pi^-}$ (MeV)
\\ \hline
a)&1299 & 300 & 0.67 $\times 10^{-10}$ \\
b)&1299 & 380 & 0.58 $\times 10^{-10}$ \\
c)&1230 & 250 & 0.72 $\times 10^{-10}$ \\
d)&1230 & 400 & 0.52 $\times 10^{-10}$ \\
e)&1209 & 576 & 0.40  $\times 10^{-10}$\\
f)&1120 & 483 & 0.35  $\times 10^{-10}$\\
\hline
\hline
\end{tabular}
\end{center}
\end{table}

In the sets (a) and (b) of the Table I we use the input data of the COMPASS Collaboration \cite{Wallner18,Aghasyan18}. Two values of $\Gamma_{a_1}$ are considered: the lowest one $\Gamma_{a_1}=300\,\mbox{MeV}$, and the central one $\Gamma_{a_1}=380\,\mbox{MeV}$. The PDG averaged values (c) and (d) \cite{PDG18} are selected in the same way. The input (e) is taken in accord with the results of the JPAC group \cite{Mikhasenko18}. In the set (f) the output of the analyses \cite{Pich10} is used.    

To summarize the above, it should be noted that, the ground states contribution, where the $a_1(1260)$ exchange dominates, is too low to explain the experimental data on $\Gamma_{\tau\to\nu_\tau\rho^0\pi^-}$. The best estimate here is given by the  set (c), but even this prediction of the NJL model is slightly below the lower boundary of the experimental value (\ref{exp}). Therefore, one should take into account the exited states which also contribute to the $\tau\to\nu_\tau \rho^0 \pi^-$ decay amplitude in leading $1/N_c$ order.

\subsection{Exited states contribution}
Let us turn now to the study of the contributions arising from the exited $\pi (1300)=\pi'$ and $a_1(1640)=a'_1$ states to clarify their role in the decay $\tau \to\nu_\tau\rho^0\pi^-$. 

The characteristics of $\pi(1300)$ quoted by the PDG are $m_{\pi'}=1300\pm 100\,\mbox{MeV}$, and $\Gamma_{\pi'}=200-600\,\mbox{MeV}$ \cite{PDG18}. The impact of this state on the $\tau\to\nu_\tau \rho^0 \pi^-$ decay width is controlled by the parameter $\beta_{\pi'}$. In accord with the PCAC relation, one can expect that $|\beta_{\pi'}|\sim (m_\pi /m_{\pi'})^2 = 0.01$ \cite{Weiss97}. This is too small to have an appreciable impact on the $\tau\to \nu_\tau\rho^0\pi^-$ decay width. Hence, the only contribution which may affect the description presented in Table \ref{scase} is the contribution of the exited $a_1(1640)$ state. 

The PDG lists the $a_1(1640)$ as "omitted from summary table", nonetheless they give the average world values $m_{a_1'}=1654\pm 19\,\mbox{MeV}$, and $\Gamma_{a_1'}=240\pm 27\,\mbox{MeV}$. On top of that, the COMPASS Collaboration has recently reported on the Breit-Wigner $a_1'$-resonance parameters: $m_{a'_1}= 1700^{+35}_{-130}\,\mbox{MeV}$, and  $\Gamma_{a'_1}=510^{+170}_{-90}\,\mbox{MeV}$ \cite{Aghasyan18}. 

It is worth mentioning that the value of the parameter $\beta_{a'_1}$ is not so strongly suppressed as $\beta_{\pi'}$. This follows from the crude estimate $|\beta_{a'_1}|\sim (m_{a_1} /m_{a'_1})^2 = 0.56$. Therefore one can expect that mixing (\ref{sub}) gives a visible effect [let us note, that a similar estimation made for the parameter $\beta_{\rho'}$ considered in \cite{Pich10} in the context of an effective description of the role of the $\rho' =\rho(1450)$ exited state of $\rho(770)$ gives $|\beta_{\rho'}|\sim (m_\rho /m_{\rho'})^2 =0.28$, in harmony with their result $\beta_{\rho'}=-0.25$, obtained by fitting experimental data]. In the following, the free parameter $\beta_{a'_1}$ will be fixed in accord with our estimate $\beta_{a'_1}=-0.56$ above. Notice the increase of the impact of the ground state $a_1(1260)$ due to the factor $1/(1+\beta_{a'_1})$ in (\ref{sub}). Again, the similar effect took place for the ground $\rho(770)$ state contribution when the exited state $\rho' =\rho(1450)$ had been taken into account \cite{Pich10}. 
 
 Our goal now is to show that the known experimental data allow for a meaningful evaluation of the impact of the $a_1'$-resonance propagator on the $\tau \to\nu_\tau\rho^0\pi^-$ decay. To this end, we consider the sets of experimentally known characteristics of $a_1$ and $a_1'$ resonances. The results of such numerical calculations are collected in Table \ref{fcase}.
 
 In the sets (a) and (b), the data of the COMPASS Collaboration are considered \cite{Aghasyan18}. Notice that COMPASS has performed the so far most advanced partial-wave analysis of diffractively produced $\pi^+\pi^-\pi^-$ final states, using the isobar model. That has allowed them, in particular, to determine mass and width of $a_1$ and $a_1'$ resonances with high confidence. Their interpretation of $a_1'$ as a first radial excitation of $a_1$ is in line with our theoretical consideration. 
 
\begin{figure}
\resizebox{0.45\textwidth}{!}{%
 \includegraphics{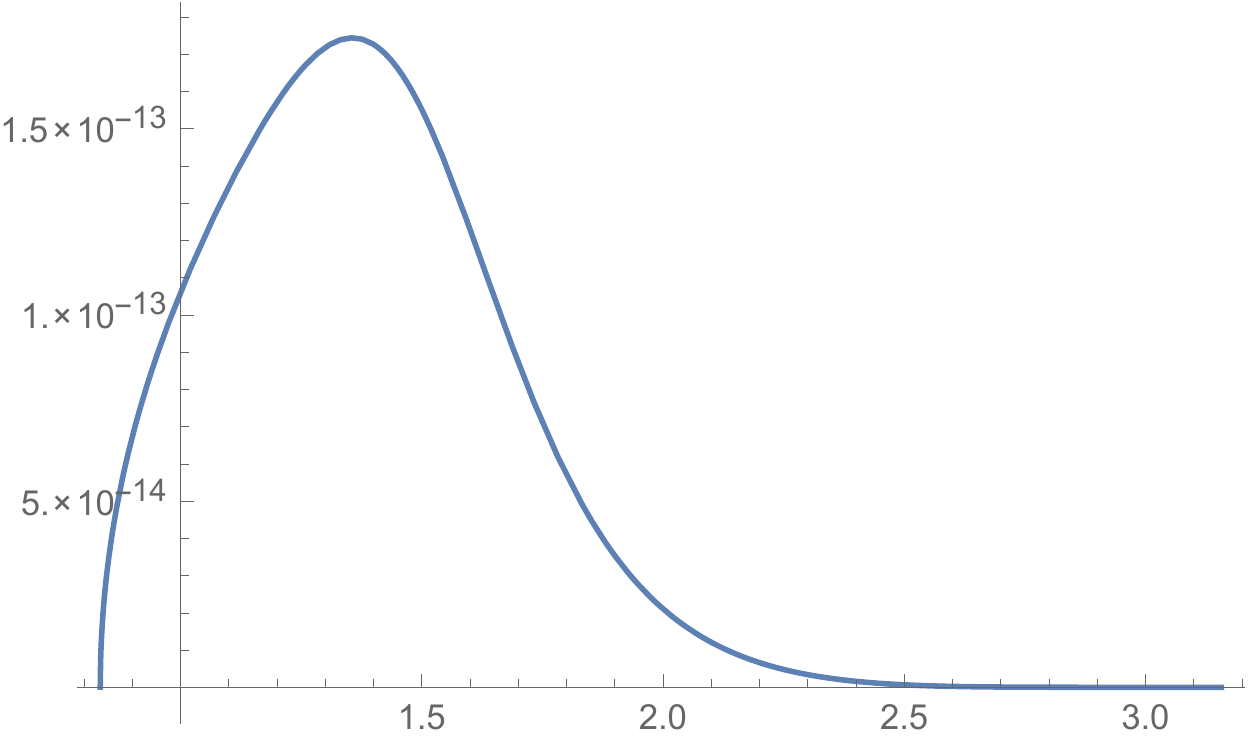}}
\caption{The predicted spectral distribution (\ref{spf}) (in GeV$^{-1}$ on the ordinate) as a function of $s$ (in GeV$^2$ on the abscissa). The parameters correspond to the set (b) in Table \ref{fcase}. }
\label{fig3}      
\end{figure}

 In Fig.\ref{fig3} we show the typical behaviour of the spectral function (\ref{spf}) for the case (b), which agrees well both with the experimental value (\ref{exp}) and with the JPAC estimate (\ref{JPAC}). The $a_1(1640)$ resonance contributes mostly through its interference with $a_1(1260)$. The distractive interference suppresses the $a_1$-exchange contribution  $\Gamma^{(a_1)}_{\tau\to\nu_\tau\rho^0\pi^-}$ on $16\%$.  
   
The sets (c) and (d) are based on the PDG averaged values: $m_{a_1}=1230\pm 40\,\mbox{MeV}$, $\Gamma_{a_1}=250-600\,\mbox{MeV}$, and $m_{a_1'}=1654\pm 19\,\mbox{MeV}$, $\Gamma_{a_1'}=240\pm 27\,\mbox{MeV}$ \cite{PDG18}. The width of $a_1$ shows large uncertainties, but large values for the width are known to be ruled out by COMPASS measurements. Thus, in our estimates, we use two values $\Gamma_{a_1}=250\,\mbox{MeV}$ and $400\,\mbox{MeV}$. The latter value is more preferable. The distractive interference suppresses the $a_1$-exchange contribution $\Gamma^{(a_1)}_{\tau\to \nu_\tau\rho^0\pi^-}$ on $20\%$.    
 
In the set (e) we use the $a_1$ parameters extracted by the JPAC group $m_{a_1}=1209\pm 4^{+12}_{-9}\,\mbox{MeV}$, and $\Gamma_{a_1}=576\pm 11^{+80}_{-20}\,\mbox{MeV}$ \cite{Mikhasenko18}. In the set (f) the data of the theoretical fit for the ground state of $a_1$ are considered \cite{Pich10}. In both cases the characteristics of $a'_1$ are taken from the PDG \cite{PDG18}.

\begin{table}[tb]
\caption{The effect of the exited axial-vector $a_1(1640)$ state on the $\tau\to\nu_\tau\rho^0\pi^-$ decay width obtained in the NJL model with the use of Eq.(\ref{sub}) and Eq.(\ref{Rwidth}), where $\beta_{\pi'}=0$ and $\beta_{a'_1}=-0.56$.  The ground state contribution $\Gamma^{(a_1)}_{\tau\to\nu_\tau\rho^0\pi^-}$ is shown for comparison. The masses and widths are given in MeV.}
\label{fcase}
\begin{center}
\begin{tabular}{ccccccccc}
\hline
\hline
    set &
    $m_{a_1}$  & 
    $\Gamma_{a_1}$ &
    $m$ &
    $g_A$ &
    $m_{a'_1}$  &
    $\Gamma_{a'_1}$ &
    $\Gamma^{(a_1)}_{\tau\to\nu_\tau\rho^0\pi^-}$   & 
    $\Gamma_{\tau\to\nu_\tau \rho^0\pi^-}$ 
\\ \hline
a) &1299 & 300  & 426   & 0.356    & 1700   & 510 & $1.95 \times 10^{-10}$ &  $1.35 \times 10^{-10}$ \\
b) &1299 & 380  & 426   & 0.356    & 1700   & 510 & $1.56 \times 10^{-10}$ & $1.31 \times 10^{-10}$ \\
c) &1230 & 250  & 390   & 0.397    &  1654  &  240 & $2.74 \times 10^{-10}$ &  $2.22 \times 10^{-10}$\\
d) &1230 & 400  & 390   & 0.397    &  1654  &  240 & $1.61 \times 10^{-10}$ &  $1.28 \times 10^{-10}$\\
e) &1209 & 576  & 379   & 0.411    &  1654  &  240 & $1.12 \times 10^{-10}$ &  $0.83 \times 10^{-10}$\\
f) &1120  & 483  & 330   & 0.479    & 1654  &  240 & $1.17 \times 10^{-10}$ &  $0.99 \times 10^{-10}$\\
\hline
\hline
\end{tabular}
\end{center}
\end{table}

Let us summarize the results presented in Table \ref{fcase} and Fig. \ref{fig3}. 

1) The $a_1(1260)$ resonance dominates the $\tau\to\nu_\tau\rho^0\pi^-$ process, while the $a_1(1640)$ contributes less than the $20\%$.

2) The contribution of the $\pi(1300)$ resonance to the $\tau\to\nu_\tau\rho^0\pi^-$ decay is negligible. This is a direct consequence of the PCAC relation. In particular, the value $\beta_{\pi'}=0$ in Table \ref{fcase} can be replaced by $\beta_{\pi'}=-(m_\pi /m_{\pi'})^2=-0.01$ without a noticeable effect. To have a noticeable effect the value of $\beta_{\pi'}$ should be about $\beta_{\pi'} =-0.4$. At this stage, however we do not see any valid theoretical reason why $|\beta_{\pi'}|$ would be so large.     

3) The comparison of Tables \ref{scase} and \ref{fcase} shows that the inclusion of the excited axial-vector $a_1(1640)$ state is a necessary element for the successful description of the $\tau\to\nu_\tau\rho^0\pi^-$ decay width. A reason for the improvement is contained in the substitutions (\ref{sub}) and (\ref{Rwidth}) which increase substantially the contribution of the $a_1$ ground state, although this growth is partly suppressed due to a destructive interference with the exited $a_1'$ state. This our conclusion agrees with the CLEO Collaboration result \cite{Asner99}. Their studies show that adding of the $a'_1$ term into the Breit-Wigner function improves significantly the agreement with the $\tau\to\nu_\tau 3\pi$ data. 
 
4) The set c) overestimates the $\tau \to \nu_\tau \rho^0\pi^-$ decay rate. This is a consequence of a very low $a_1$-resonance width $\Gamma_{a_1}= 250\,\mbox{MeV}$. The other sets with larger values of $\Gamma_{a_1}$ are in agreement with the experimental value (\ref{exp}). Apparently, this points out that the value $\Gamma_{a_1}\simeq 400\,\mbox{MeV}$ is preferable one. This observation is consistent with the determinations from COMPASS.  

5) The spectral distribution shown in Fig. \ref{fig3} is a prediction of the NJL model. It is assumed that this result would be checked in the study of the tau decay into three pions and neutrino, where events with pion pairs over the $\rho (770)$ mass would be selected.  

\section{Conclusions}
\label{concl}

The purpose of this paper has been to describe the $\tau\to\nu_\tau \rho^0\pi^-$ decay by using an extended $U(2)_L\times U(2)_R$ chiral symmetric NJL model with spin-0 and spin-1 four-quark interactions. The channel $\tau\to\nu_\tau \rho^0\pi^-\to \nu_\tau \pi^-\pi^-\pi^+$ dominates the tau decay into three pions and neutrino. That explains our interest to the problem. 

We have used the covariant approach \cite{OsipovHZ18} to describe the weak interactions of mesons in leading order in $1/N_c$ and derivative expansions. It has been shown that in this approximation the axial-vector current is dominated by the $a_1(1260)$ and the pion exchanges. However, we have found that the $\tau\to\nu_\tau \rho^0\pi^-$ decay width is too low if the physical value of $\Gamma_{a_1}$ is considered.  

The contributions of the first radial excitations of the pion and the $a_1$ states have been taken into account to improve the description. For that we supplemented the regular $\pi$ and $a_1$ propagators with new terms corresponding to the propagators of excited $\pi (1300)$ and $a_1(1640)$ states. Our treatment of these excitations is similar to the successful description of the ground $\rho (770)$ and excited $\rho (1450)$ vector resonances in \cite{Pich10}. The momentum dependent off shell widths of all resonances have been approximated by the functions introduced in paper \cite{Geffen78}. This procedure can be further elaborated as soon as the new more precise experimental data will be reported on the $\tau\to\nu_\tau \rho^0\pi^-$ decay.

As a result, we obtain that the contribution of the $\pi (1300)$ resonance is negligible, and conclude that the channel $\tau\to\nu_\tau \rho^0\pi^-\to \nu_\tau \pi^-\pi^-\pi^+$ is a source of sufficiently clear information on $a_1(1260)$ and $a_1(1640)$ states. The $a_1(1260)$ resonance dominates the intermediate process, while the $a_1(1640)$ contributes less than $20\%$. In Table \ref{fcase}, we present our estimations for the decay width $\Gamma (\tau\to\nu_\tau \rho^0\pi^-)$ which correspond to the different input values of $a_1$ and $a_1'$ characteristics. The spectral distribution shown in Fig. \ref{fig3} can be used for comparison with the data, as soon as those would be available. 

Our result indicates on the important role which the $a_1(1640)$ state plays in the theoretical description of this $\tau\to\nu_\tau \rho^0\pi^-$ decay. It means, in particular, that one should carefully estimate its contribution and role in the $\tau\to\nu_\tau \pi^-\pi^-\pi^+$ decay. This will be done somewhere else. The results obtained here could be useful for such studies.

\section*{Acknowledgments}
The author have benefited from innumerable discussions with M. K. Volkov and B. Hiller. A conversation with H. G. Dosch on the subject of PCAC relation in the presence of the radially excited states $\pi (1300)$ and $a_1(1640)$ is gratefully acknowledged. I would like also to thank P. Roig for useful correspondence. This paper was completed at the Institute of Modern Physics of the Chinese Academy of Sciences in Lanzhou and I would like to thank P. M. Zhang and L. Zou for their warm hospitality and support. I would like to acknowledge networking support by the COST Action CA16201.

\end{document}